\documentclass[aps,prd,a4paper,nofootinbib,
preprintnumbers,superscriptaddress,showpacs,showkeys,twocolumn]{revtex4-2}

\usepackage{amsmath}
\usepackage{amssymb}
\usepackage{bm}
\usepackage{graphicx}
\usepackage{color}
\newcommand{\be}{\begin{equation}}
	\newcommand{\ee}{\end{equation}}
\newcommand{\ba}{\begin{eqnarray}}
	\newcommand{\ea}{\end{eqnarray}}

\begin{document}
	
	\title{Thermalization and isotropization of heavy quarks in
		a non-Markovian medium in
		high-energy nuclear collisions}
	
	\author{Pooja}
	\affiliation{School of Physical Sciences, Indian Institute of Technology Goa, Ponda-403401, Goa, India}

	\author{Santosh K. Das}
	\affiliation{School of Physical Sciences, Indian Institute of Technology Goa, Ponda-403401, Goa, India}
	
	\author{Vincenzo Greco} 
	\affiliation{Department of Physics and Astronomy "Ettore Majorana", University of Catania, Via S. Sofia 64, I-95123 Catania, Italy}
	\affiliation{INFN-Laboratori Nazionali del Sud, Via S. Sofia 62, I-95123 Catania, Italy}
	
	\author{Marco Ruggieri}\email{marco.ruggieri@dfa.unict.it}
	\affiliation{Department of Physics and Astronomy "Ettore Majorana", University of Catania, Via S. Sofia 64, I-95123 Catania, Italy}\affiliation{INFN-Sezione di Catania, Via S. Sofia 64, I-95123 Catania, Italy}

	% ************************* Start : Abstract ******************************* %
\begin{abstract}
We study the 
isotropization and thermalization of heavy quarks in
a non-Markovian medium in 
high energy nuclear collisions. In particular, we 
analyze the case of a non-stationary medium with
a noise whose time-correlator
decays as a power law (heavy tailed noise). We assume the 
correlations decay with an exponent $\beta-1$,
		$0\leq\beta<1$;
		we treat $\beta$ as a free parameter.
		We analyze 
		the effect of memory on the
		thermalization and isotropization of 
		heavy quarks in the medium via a generalized 
		Langevin equation. 
		In general, we find that memory
		slows down the dynamics of heavy quarks;
moreover, thermalization and isotropization happen
on the same time scale once a realistic initialization is 
considered.		
		We also find that 
		while the effect
		on charm quarks can be relevant, beauty quarks
		are hardly affected by memory in the 
		quark-gluon plasma phase. Finally, we comment on 
		the effect of memory on the estimate of $D_s$
		of charm and beauty.

	\end{abstract}
	% ************************* End : Abstract ******************************* %

	% ~~~~~~~~~~~~~~~~~~~~~~~~~~~~~~~~~~~~~~~~~~~~~~~~~~~~~~~~~~~~~~~~~~~~~~~~ %
	
	\maketitle
	% ~~~~~~~~~~~~~~~~~~~~~~~~~~~~~~~~~~~~~~~~~~~~~~~~~~~~~~~~~~~~~~~~~~~~~~~~ %

	% ######################## Start : Introduction ########################  %
	\section{Introduction}
	
	The ultra-relativistic collision experiments performed at RHIC and LHC confirm the existence of a locally equilibrated state of free quarks and gluons known as quark-gluon plasma (QGP)~\cite{Shuryak:2004cy,Jacak:2012dx, Das:2022lqh}. The formation of QGP is a consequence of pre-equilibrium effects that happen to occur just after the collision of high-energy nuclei. It is assumed that immediately after the collision, the dynamics is
	that of dense color-electric
	and color-magnetic fields, namely the Glasma~\cite {Kovner:1995ja,Kovner:1995ts, Lappi:2006fp} and later, it decays to a system of 
strongly interacting quarks and gluons	
	which we call the quark-gluon plasma (QGP). 
The expansion of QGP continues to unbound states of hadrons after which chemical and kinetic freeze-outs take place and the particles fly towards  the detectors~\cite{Cleymans:1998fq,Shuryak:1999zh}.
	
	The dynamics of QGP is governed by light quarks and gluons, along with few heavy quarks. The heavy quarks (HQs)~\cite{Prino:2016cni,Andronic:2015wma,Rapp:2018qla,Aarts:2016hap,Cao:2018ews,Dong:2019unq,Xu:2018gux,Gossiaux:2008jv,GolamMustafa:1997id,Uphoff:2011ad,Song:2015sfa,Cao:2016gvr,Plumari:2017ntm,Das:2016llg} are formed
	 very early in the collision experiments,
	 and  are considered to be effective probes to study the evolution of QGP and Glasma as well. Their large masses,
	 $m$, lead to their early production in the medium, 
$\tau_\mathrm{prod}\sim 1/m$,
	 hence they can witness the entire evolution of the system from the very beginning of the medium formation. 
	 HQs approximately
	 undergo a Brownian motion in a medium of light quarks and 
gluons~\cite{Das:2013kea} 
and their dynamics can be studied within the framework of Langevin and Fokker-Planck equations where the interaction is taken care of in terms of diffusion and drag coefficients~\cite{rappv2,rappprl,Das:2010tj,Alberico:2011zy,Lang:2012cx,He:2012df,He:2013zua,Das:2013kea,Cao:2015hia,Das:2016cwd,Xu:2017obm,Katz:2019fkc,Li:2020umn}.

	In the vast majority of the studies related to HQs in QGP,
	the effects of memory are ignored~\cite{Uphoff:2011ad,Song:2015sfa,Cao:2016gvr,Plumari:2017ntm,rappv2, rappprl,
		Gossiaux:2008jv,Das:2010tj,Alberico:2011zy,Lang:2012cx,He:2012df,He:2013zua,Das:2013kea,Cao:2015hia,Das:2016cwd,Xu:2017obm,Katz:2019fkc,Li:2020umn,Das:2016llg,Nahrgang:2014vza,Song:2015ykw,Das:2017dsh}. However, it is plausible to assume that correlations of the 
	forces that act on HQs within the whole evolution of the
	fireball exist, in particular when the system approaches
	the phase transition;
	moreover, these correlations certainly exist
	in the early stage, due to the arrangement of the strong
	gluon fields in the form of correlated domains in the 
	transverse plane~\cite{Liu:2019lac,Liu:2020cpj,Khowal:2021zoo}.
	Several recent studies~\cite{Schmidt:2014zpa,Schuller:2019ega,Kapusta:2011gt,Murase:2016rhl,Kapusta:2012zb,Kapusta:2017hfi,Schenke:2006uh,Hammelmann:2018ath,Ruggieri:2019zos,Chen:2023cwv, oliveira2019anomalous} indicate that the memory effect plays an important role. 
	It was shown in~\cite{Ruggieri:2022kxv}
	that even when the memory time is of the order of $1$ fm/c,
	this might have an impact on observables related to HQs,
	for example the nuclear modification factor.
	In~\cite{Ruggieri:2022kxv}, a specific form of the noise
	correlations was assumed, namely an exponential one,
	which is characterized by a memory time,
	$\tau$, that sets up the scale
	for the decay of the correlations;
there it was also shown that the evolution of HQs is 
	unaffected by memory when their evolution time is much larger than $\tau$.
	
	The purpose of our study is to extend the study
	of~\cite{Ruggieri:2022kxv} to the case in which the
	correlations decay with a power law. 
	Such power laws correlations appear in different contexts
	in many areas of Physics and Chemistry~\cite{metzler2000random,chen2010anomalous,sokolov2005diffusion,Vlahos2008NormalAA}
	and usually appear in presence of strong correlations
	in the medium: it is therefore
	worth studying their potential effects on HQs in QGP as well.
	Differently from
	the case study of~\cite{Ruggieri:2022kxv}, 
	we find that HQs can be affected
	by the presence of correlations even in the late stage
	of the evolution. In particular,
	in this study we analyze momentum isotropization
	and thermalization of HQs in a QGP bath, 
	and quantify the effects of memory on these processes.
	We can anticipate our results here, namely that
	not only the presence of memory delays both thermalization and
	isotropization, but also that the specific form of the 
	noise correlations affect the late time evolution of the
	system. We estimate the thermalization time of HQs in 
	a bath with memory, both for charm and beauty quarks,
	finding that charm quarks are more affected than beauty quarks.
	Finally, from the estimate of the
	thermalization time we evaluate the effect of memory
	on the spatial diffusion coefficient $D_s$.

	The plan of the article is as follows. In Section II we present
	the formalism and explain how the power law processes
	are implemented. In Section III we present our results.
	Finally, in Section IV we summarize our conclusions.

	% ######################## End : Introduction ########################  %

	% ++++++++++++++++++++++++++ Start : Formalism ++++++++++++++++++++++++++ %
	
	\section{Formalism\label{sec:formalism}}
	\subsection{Noise with power law memory\label{subsec:noise_h}}
	In this section, we discuss the method to implement 
	a longtail noise whose correlations
	decay as a power law. 
	We introduce the process
	\begin{equation}
		h(t) =\sqrt{\kappa} \frac{\sqrt{\beta}}{\tau^\beta}
		\int_0^t (t-u)^{\beta-1}\xi(u) du,
		\label{eq:Caputo_111}
	\end{equation}
	where  $0<\beta<1$; $\xi$ is a standard Gaussian noise with zero average and time correlations given by
	\begin{equation}
		\langle \xi(t_1) \xi(t_2)\rangle = \tau\delta(t_1-t_2).
		\label{eq:Caputo_corr_222}
	\end{equation}
	We note that 
	Eq.~\eqref{eq:Caputo_111} is
	proportional to
	the Riemann-Liouville fractional 
	integral of $\xi$ of order $\beta$:
	in fact, besides the overall constant, the process $h$
	in~\eqref{eq:Caputo_111} corresponds to the formal solution of the fractional Langevin equation
	$D^\beta h=\xi$ where $D^\beta$ denotes the fractional
	derivative of order $\beta$. 
	We introduce the free parameter $\tau$, with dimension of time, 
	so $\tau^{-\beta}$ in front of the integral in 
	Eq.~\eqref{eq:Caputo_111} balances the dimension 
	of the integral itself giving a dimensionless $h$, 
	and $\tau$ balances the dimension of the $\delta-$function in Eq.~\eqref{eq:Caputo_corr_222} to give a dimensionless $\xi$.
	We will show later that $\tau$ sets the time scale
	in the decay of the correlations of $h$,
	while $\beta$ fixes the power law at which
	correlations decay.
	The overall $ \sqrt{\beta}$ is added for later 
	convenience, to simplify the expressions 
	of the correlator of the noise and of the
	momentum broadening in the purely diffusive motion.
	Finally, $ \kappa=1/8.44$
	is introduced to reproduce the 
	momentum spreading 
	of the memoryless processes 
	in the limit
	$\beta\rightarrow0$, see section~\ref{subsec:pure_diffusion_Langevin} and Fig.~\ref{Fig:mombrodif}.

	The definition~\eqref{eq:Caputo_111}
	has to be understood in the sense of 
	the It\={o} calculus,
	namely, it corresponds to 
	\begin{equation}
		h(t_N) =  \frac{\sqrt{\beta\kappa}}{\tau^\beta}
		\Delta t\sum_{i=0}^{N-1}
		(t_N-t_i)^{\beta-1}\xi(t_i),
		\label{eq:Caputo_111_bis}
	\end{equation}
	where we assume that the
	process happens from $t_0=t_\mathrm{initialization}$ to $t_N=t$ in $N$ 
	time steps, 
	each of width $\Delta t=(t_N-t_0)/N$, hence
	$t_i=t_0+i\Delta t$.
	By virtue of Eq.~\eqref{eq:Caputo_corr_222} it is
	easy to prove that 
	the time correlations of $h$ are given by
	\begin{equation}
		\langle h(t_1) h(t_2)\rangle =
		\kappa	\tau^{-2\beta+1} \beta
		\int_0^{t_\mathrm{min}}(t_1-u)^{\beta-1}
		(t_2-u)^{\beta-1}du,
		\label{eq:Caputo_corr_111}
	\end{equation}
	where $t_\mathrm{min}=\mathrm{min}(t_1,t_2)$.
	The integral on the right-hand side
	of Eq.~\eqref{eq:Caputo_corr_111} can be 
	expressed in terms of an incomplete Euler
	beta-function, namely
	\begin{equation}
		\langle h(t_1) h(t_2)\rangle =
		\kappa	\tau^{-2\beta+1} \beta 
		(t_1-t_2)^{2\beta-1}
		(-1)^{-\beta}B_Z(\beta,\beta),
		\label{eq:Caputo_corr_111exact}
	\end{equation}
	with $Z=-t_2/(t_1-t_2)$ and
	\begin{equation}
		B_Z(x,y)=\int_0^Z u^{x-1}(1-u)^{y-1} du.
	\end{equation}
	It is easy to see that for $t_1\gg t_2$ the
	correlator~\eqref{eq:Caputo_corr_111}
	decays as a power law. In fact, 
	in the limit $t_1\gg t_2$ 
	the factor $(t_1-u)$ in the integral in 
	Eq.~\eqref{eq:Caputo_corr_111} can be replaced by $t_1$, 
	so that 
	\begin{eqnarray}
		\langle h(t_1) h(t_2)\rangle &\approx&
		\kappa\tau^{-2\beta+1}
		\beta
		t_1^{\beta-1}\int_0^{t_2} 
		(t_2-u)^{\beta-1}du\nonumber\\
		&=& \kappa \left(\frac{t_1}{\tau}\right)^{\beta-1} \left(\frac{t_2}{\tau}\right)^{\beta}.
		\label{eq:Caputo_corr_111_asy_2}
	\end{eqnarray}
	We note that $\beta>0$ is enough to ensure
	the convergence of the integral above.
	Thus, 
	for fixed $t_2$, the correlations of the process \eqref{eq:Caputo_111} decay for $t_1\gg t_2$ with the power law $1/t_1^{1-\beta}
	\approx 1/( t_1-t_2)^{1-\beta}$:
	the smaller $\beta$ implies the faster decay of correlations.
	Since time correlations in $h$ exist, we say that
	this process has a memory; moreover, since the
	time correlations decay with a power law, we say that
	the process is characterized by a longtail memory,
	to distinguish it from the processes studied before
	in which the correlations are damped exponentially.
	Sometimes these processes are called heavy tailed processes,
	to emphasize that the correlations of the noise do not decay
	exponentially with time.
	We also note that the correlator does not depend on
	$t_1-t_2$ but on $t_1$ and $t_2$ separately. 
	This will lead to a non-stationary Langevin equation
	in the next section.

	The $\delta-$function in Eq.~\eqref{eq:Caputo_corr_222} has to be understood as 
	$\delta_{t_1,t_2}/\Delta t$.
	Consequently, it is convenient to rescale $\xi$ as
	\begin{equation}
		\xi(t) = \sqrt{\frac{\tau}{\Delta t}}\zeta(t),
		\label{eq:scaling_1_ooo}
	\end{equation}
	so that $\zeta$ is generated at each time step according to
	\begin{equation}
		\langle \zeta(t_1) \zeta(t_2)\rangle = \delta_{t_1,t_2}.
		\label{eq:Caputo_corr_222_bis}
	\end{equation}
	By virtue of the rescaling~\eqref{eq:scaling_1_ooo}, we can rewrite Eq.~\eqref{eq:Caputo_111_bis} as
	\begin{equation}
		h(t_N) =   \tau^{ -\beta+1/2}  
		\sqrt{\kappa\beta\Delta t}\sum_{i=0}^{N-1}
		(t_N-t_i)^{\beta-1}\zeta(t_i).
		\label{eq:Caputo_111_ter}
	\end{equation}
	where $\zeta(t)$ corresponds to white noise with variance equal to one.
	% ~~~~~~~~~~~~~~~~~~~~~~~~~~~~~~~~~~~~~~~~~~~~~~~~~~~~~~~~~~~~~~~~~~

	\subsection{Generalized Langevin equation}
	In our study, we couple
	the noise $h$ discussed in the previous subsection to heavy quarks
	via a generalized Langevin equation: 
	for simplicity we present the formulation
	of a one-dimensional motion, 
	while actual numerical calculations
	will be run for the three-dimensional case. 
	
The  Langevin equation for momentum $p$ 
	reads \cite{batta2008,vogt2005,hernandez1,hernandez2,hernandez3,hernandez4,popov2007,kawai2011,meyer2017,book1,book2}
	\begin{equation}
		\frac{dp(t)}{dt}=-\int_0^t dt^\prime~\gamma(t,t^\prime)p(t^\prime)+\eta(t),
		\label{eq:L1}
	\end{equation}
	where the integral term  represents the dissipative force and $\eta(t)$ is the thermal noise in a 
	bath at the temperature $T$.
	We note that the we assume the dissipative kernel,
	$\gamma$, is a function of both $t$ and $t^\prime$:
	this is called an irreversible generalized Langevin equation~\cite{Kubo:1966gt,batta2008,vogt2005,hernandez1,hernandez2,hernandez3,hernandez4,popov2007,kawai2011,meyer2017,book1,book2},
	as it generalizes the Langevin equation to the motion
	of probes in 
nonstationary baths that are characterized,
for example, by time and/or
	space changes in the bath temperature.
	In heavy ion collisions, modeling  a $\gamma=
	\gamma(t,t^\prime)$  
	could be relevant,
	because the medium evolution
is  not invariant under time translations
even though the system
	is locally in thermal equilibrium.
	
	Following the notation of~\cite{Ruggieri:2022kxv}
	we
	assume that $\eta(t)$ in \eqref{eq:L1} satisfies 
	\begin{eqnarray}
		\langle\eta(t)\rangle &=&0,\\
		\langle\eta(t_1)\eta(t_2)\rangle &=&  2{\cal D}
		\frac{ g(t_1,t_2)}{2\tau},
		\label{eq:sbd}
	\end{eqnarray}
	where $\cal D$ is the diffusion coefficient and $g$ is a dimensionless function that defines the correlation of the noise; 
	the factor $1/2\tau$ 
	in Eq.~\eqref{eq:sbd}
	is introduced
	to balance the dimension of
	$\mathcal{D}$ so that
	the dimensions of the left and the right-hand sides
	of the
	equation match.
	In the case of a Markov process $g(t)/2\tau=\delta(t)$.
	We put
	\begin{equation}
		\eta(t)=  \sqrt{\frac{{\cal D}}{\tau}}h(t),	\label{eq:recale_eta}
	\end{equation}
	where $h$ is the process
	introduced in section \ref{subsec:noise_h}.
	Hence, 
	\begin{equation}
		\langle\eta(t_1)\eta(t_2)\rangle =  \frac{{\cal D}}{\tau} \langle h(t_1) h(t_2)\rangle,
		\label{eq:sbd_due}
	\end{equation}
	where the correlator on the right-hand side is given by Eq.~\eqref{eq:Caputo_corr_111}. 
	The comparison with Eq.~\eqref{eq:sbd} gives
	\begin{eqnarray}
		g(t_1,t_2) 	&=& \langle 	h(t_1) h(t_2)\rangle,
		\label{eq:daref789}
	\end{eqnarray}

	In terms of $h$, the Langevin equation~\eqref{eq:L1} becomes
	\begin{equation}
		\frac{dp(t)}{dt}=-\int_0^t dt^\prime~\gamma(t,t^\prime)p(t^\prime) + \sqrt{\frac{\mathcal{D}}{\tau }}h(t).
		\label{eq:L1b}
	\end{equation}
	The time-discretized version of this equation reads
	\begin{equation}
		\Delta p = -\Delta t\int_0^t dt^\prime~\gamma(t,t^\prime)p(t^\prime) + \sqrt{\frac{\mathcal{D}}{\tau }}h(t)\Delta t.
		\label{eq:L1bis}
	\end{equation}
	The dissipative term in Eq.~\eqref{eq:L1bis} has to be
	understood as a It\=o integral; we can then 
	rewrite Eq.~\eqref{eq:L1bis} as
	\begin{eqnarray}
		p(t_N) &=& p(t_{N-1}) 
		-\Delta t\sum_{k=0}^{N-1}
		~\gamma(t_N,t_k)p(t_k) \Delta t
		\nonumber\\
		&&	+~\sqrt{\frac{\mathcal{D}}{\tau }}h(t_N)\Delta t,
		\label{eq:L1bis3}
	\end{eqnarray}
	with $t_0=t_\mathrm{initialization}$, $t_N = t$ and $t_{N-1}=t_N-\Delta t$.

	We define the
	kernel of the dissipative force
	as
	\begin{equation}
		\gamma(t_N,t_k)= \frac{{\cal D}}{E(t_k) T}
		\frac{\langle h(t_N) h(t_k)\rangle}{\tau} ,
		\label{eq:FDT_in_h}
	\end{equation} 
	where we took into account the rescaling~\eqref{eq:recale_eta};
	we put $E=\sqrt{\bm p^2 + m^2}$ and $m$ is the
	heavy quark mass. The definition~\eqref{eq:FDT_in_h} is inspired
	by the Einstein relation between the drag and the
	diffusion coefficient in a medium~\cite{Das:2013kea,Walton:1999dy,Rapp:2009my,popov2007}.
	Equations~\eqref{eq:Caputo_111_ter},~\eqref{eq:L1bis3}
	and~\eqref{eq:FDT_in_h} represent the whole
	process we implement in our study.
	
	When we compare the results of the process $p(t)$
	in Eq.~\eqref{eq:L1} with a memoryless process,
	we  replace
	\begin{equation}
		\frac{g(t_1,t_2)}{2\tau}\rightarrow\delta(t_1-t_2)
		\label{eq:repla_1}
	\end{equation}
	in Eq.~\eqref{eq:sbd};
	hence correlations of the noise in this case read
	\begin{eqnarray}
		\langle\eta(t_1)\eta(t_2)\rangle &=&  2{\cal D}
		\delta(t_1-t_2)=
		\frac{2{\cal D}}{\Delta t}\delta_{t_1,t_2}.
		\label{eq:sbd_22a}
	\end{eqnarray}
	Instead of Eq.~\eqref{eq:FDT_in_h} we then have
	\begin{equation}
		\gamma(t_N,t_k)=\frac{2{\cal D}}{E T}   \delta(t_N-t_k)
		\equiv
		2\gamma \delta(t_N-t_k),
		\label{eq:FDT_in_h_mc}
	\end{equation}
	and Eq.~\eqref{eq:L1} becomes
	\begin{equation}
		\frac{dp(t)}{dt}=- \gamma p(t)+\eta(t).
		\label{eq:L1_mc}
	\end{equation}
	Adopting the standard rescaling of the noise
	\begin{equation}
		\eta(t) = \sqrt{\frac{2\mathcal{D}}{\Delta t}}\xi,
	\end{equation}
	we can rewrite Eq.~\eqref{eq:L1_mc} as
	\begin{equation}
		\Delta p=- \gamma p\Delta t + \sqrt{2\mathcal{D}
			\Delta t}\xi,
		\label{eq:L1_mc_2}
	\end{equation}
	where $\xi$ is a gaussian noise with $\langle \xi\rangle=0$ and 
	$\langle \xi^2\rangle=1$.

	% ^^^^^^^^^^^^^^^^^^^^^^ START : New secion ^^^^^^^^^^^^^^^^^^^^^^^^^^ %

	% ++++++++++++++++++++++++++ End : Formalism ++++++++++++++++++++++++++ %

	\subsection{Purely diffusive motion and determination of
		$\kappa$
		\label{subsec:pure_diffusion_Langevin}}
	It is useful to compute the evolution 
	of $\langle p^2(t)\rangle$ in a purely diffusive, one-dimensional motion,
	\begin{equation}
		\frac{dp(t)}{dt}=\eta(t),
		\label{eq:we_have_from}
	\end{equation}
	with $\langle\eta(t)\rangle=0$ and
	\begin{equation}
		\langle\eta(t_1)\eta(t_2)\rangle =  \frac{{\cal D}}{\tau} \langle h(t_1) h(t_2)\rangle,
	\end{equation}
	in agreement with the discussion in the previous
	subsection.
	For the purpose of the present section, it is enough to
	assume that the initial momentum is $p_0=0$: if $p_0\neq0$ then $\langle p^2(t)\rangle$ should be replaced by $\langle (p(t)-p_0)^2 \rangle$.
	From~\eqref{eq:we_have_from} we have
	\begin{equation}
		\langle p^2(t)\rangle =
		\frac{{\cal D}}{\tau}\int_0^t dt_1\int_0^t dt_2\langle h(t_1) h(t_2)\rangle.
		\label{eq:eta_v_eta_u}
	\end{equation}
	In the above integral, 
	the correlator to be used is given by
	Eq.~\eqref{eq:Caputo_corr_111}.

	\begin{figure}[t!]
		\begin{center}
			\includegraphics[scale=0.23]
			{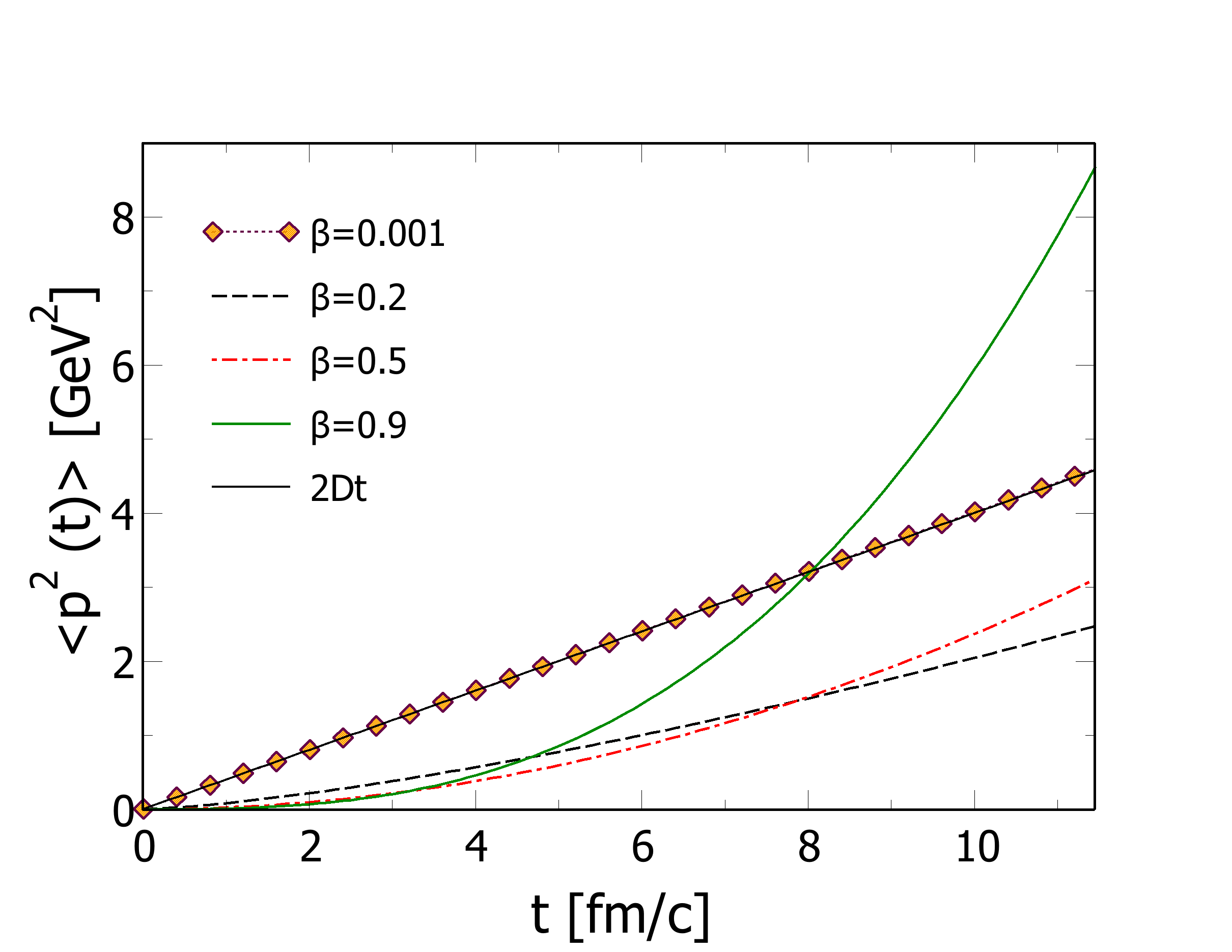}
		\end{center}
		\caption{\label{Fig:mombrodif}
			$\langle p^2\rangle$ versus time for a one-dimensional purely diffusive motion, 
			for several values of $\beta$,
			$\mathcal{D}=0.2$ GeV$^2$/fm and
			$\tau=1$ fm/c.
			We used $1/\kappa=8.44$ in
			Eq.~\eqref{eq:Caputo_corr_111}.
		For comparison we show $\langle p^2\rangle=2\mathcal{D}t$
	that would be obtained in the memoryless case.
Note that the black solid line and the orange diamonds
almost overlap.}
	\end{figure}

	In 
	Fig.~\ref{Fig:mombrodif} we plot 
	$\langle p^2\rangle$ versus time
	for several values of $\beta$,
	for $\mathcal{D}=0.2$ GeV$^2$/fm and $\tau=1$ fm/c.
	The value of $\mathcal{D}$ has been chosen 
	in agreement with the pQCD value at $T=1$ GeV.
	We note that initially, for higher $\beta$, slower diffusion
	occurs. On the other hand, for $t\gg\tau$ the trend is inverted
	and larger $\beta$ implies a faster diffusion. In particular, 
	for the smallest value of $\beta$ shown in the figure
	$\langle p^2(t)\rangle$ evolves almost linearly with time, in agreement
	with our previous discussion, while for larger $\beta$ 
	the $\langle p^2(t)\rangle$ increases with a power of time
	larger than one. 
We also note that for $\beta\rightarrow 0$ the momentum
broadening agrees with the one that would be obtained in a 
bath without memory, namely $\langle p^2\rangle=2\mathcal{D}t$:
as we show later, this result is independent of $\tau$.

	\begin{figure}[t!]
		\begin{center}
			\includegraphics[scale=0.23]
			{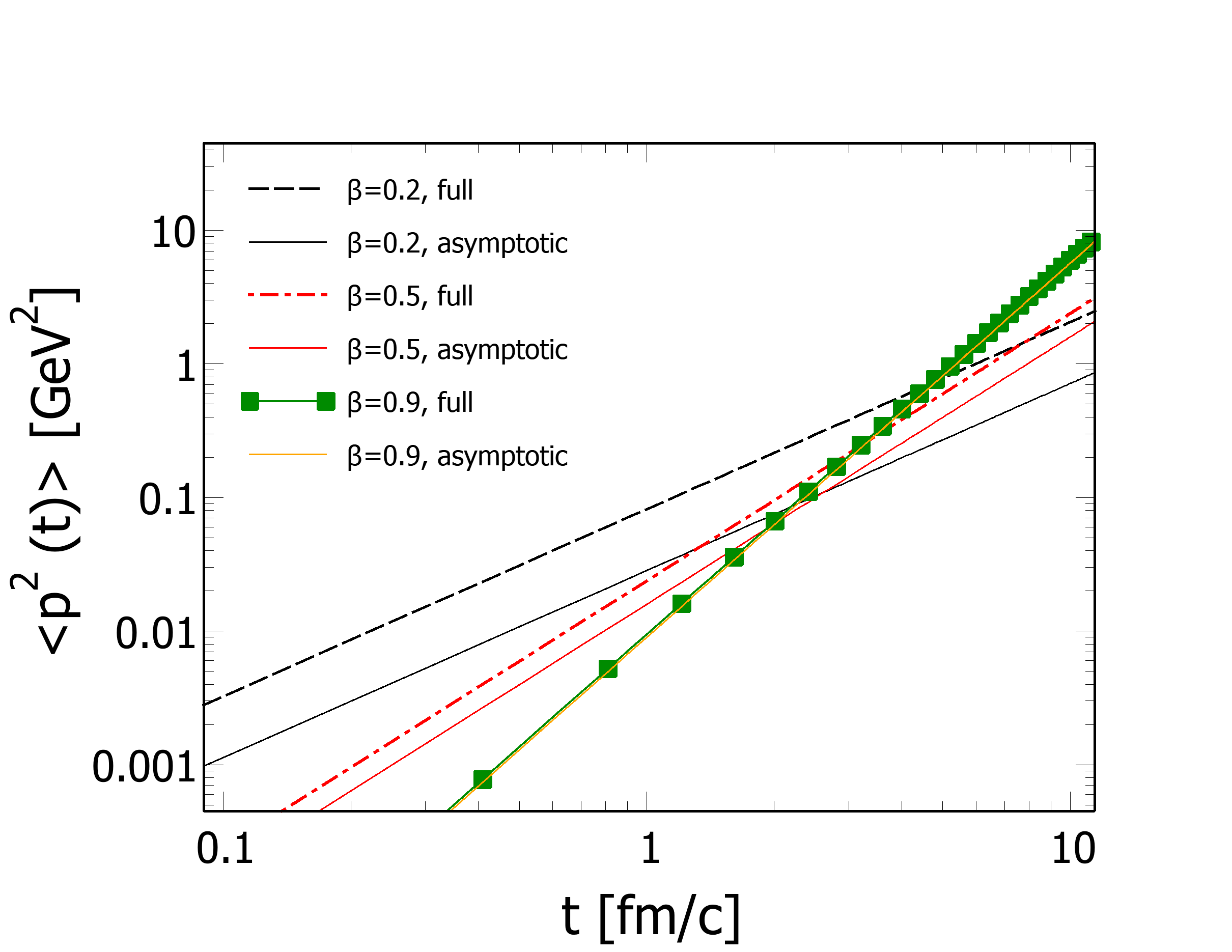}
		\end{center}
		\caption{\label{Fig:mombrodiflog}
			$\langle p^2\rangle$ versus time for a one-dimensional purely diffusive motion, 
			for three different values of $\beta$,
			$\mathcal{D}=0.2$ GeV$^2$/fm and
			$\tau=1$ fm/c. The asymptotic form corresponds to
			Eq.~\eqref{eq:eta_v_eta_u_3}. Note that the green
			squares and the thin orange line overlap.
		}
	\end{figure}  
	
	We were able to extract an approximate analytical
	expression for the time dependence of 
	$\langle p^2(t)\rangle$:
using Eq.~\eqref{eq:Caputo_corr_111_asy_2}
	in the right-hand side
	of Eq.~\eqref{eq:eta_v_eta_u}, we get
	\begin{eqnarray}
		\langle p^2(t)\rangle &=&
		\frac{\kappa{\cal D}}{\tau}
		\tau^{-2\beta+1}
		\int_0^t dt_1\int_0^{t_1} dt_2
		~t_1^{\beta-1}t_2^\beta 
		\nonumber\\
		&+&\frac{\kappa{\cal D}}{\tau}
		\tau^{-2\beta+1}
		\int_0^t dt_1\int_{t_1}^{t} dt_2
		~t_2^{\beta-1}t_1^\beta,
		\label{eq:eta_v_eta_uAA}
	\end{eqnarray}
	where we used the fact that the form~\eqref{eq:Caputo_corr_111_asy_2}
	stands for $t_1 > t_2$, so a similar result needs
	to be used for the case $t_2 > t_1$ in~\eqref{eq:eta_v_eta_u}. 
	Performing the elementary integration, we then have
	\begin{equation}
		\langle p^2(t)\rangle
		=
		\frac{2\kappa{\cal D}}
		{1+3\beta+2\beta^2}
		\left(	\frac{t }{\tau}\right)^{2\beta}t
		.
		\label{eq:eta_v_eta_u_3}
	\end{equation}
	In Fig.~\ref{Fig:mombrodiflog} we 
	compare the result~\eqref{eq:eta_v_eta_u_3} with the
	full calculation~\eqref{eq:eta_v_eta_u} 
	on a log-log scale. 
	We note that the 
	approximation~\eqref{eq:eta_v_eta_u_3} does not work
	well for small values of $\beta$, while it works
	pretty well for $\beta\approx 1$. However, it is 
	remarkable that the slopes of the approximate and
	exact solutions agree with each other in the whole
	range of $\beta$.  Hence,  while Eq.~\eqref{eq:eta_v_eta_u_3}
	cannot be used to estimate quantitatively 
	$\langle p^2(t)\rangle$ in the whole range of $\beta$,
	it is still useful to extract the time dependence
	of $\langle p^2(t)\rangle$.

	We note from Eq.~\eqref{eq:eta_v_eta_u_3}
	that parametrically $\langle p^2(t)\rangle\propto ( t/\tau)^{2\beta}t$.
	For $t\ll\tau$ the momentum diffusion
	with $\beta\rightarrow 1$ is quite slower than the
	one with $\beta\rightarrow 0$: 
	higher correlations in the noise slow down
	momentum broadening in the 
	early stage in agreement with~\cite{Ruggieri:2022kxv}.
	At later times, $t\gg\tau$, 
	Eq.~\eqref{eq:eta_v_eta_u_3} suggests that
	increasing $\beta$ results in a faster momentum
	broadening.
	This is confirmed by 
	the results shown in Fig.~\ref{Fig:mombrodif}
	in which the data with large $\beta$ overshoot
	those with small $\beta$ for $t\gg\tau$.
	We also note that the asymptotic result~\eqref{eq:eta_v_eta_u_3}
	implies that for $\beta\rightarrow0$
	the momentum broadening is independent of $\tau$.
	Hence, it seems appropriate to state that in the $\beta\rightarrow 0$
	limit, we recover the momentum
	diffusion of a memoryless process.

	%%%%%%%%%%%%%%%%%%%%%%%%% Start Results %%%%%%%%%%%%%%%%%%%%%%%%%%%%%%%%%%%%
	\section{Results\label{sec:results}}
	In this section, 
	we present our results on momentum randomization,
	isotropization and thermalization of HQs.
	For illustrative purposes,
	we firstly 
	consider a simplified initialization.
	Then, we show the impact of memory for the more realistic
	diffusion coefficients, borrowed from pQCD~\cite{comb, Svetitsky:1987gq} and 
	by a quasi-particle model (QPM)~\cite{Das:2015ana, Scardina:2017ipo}  for 
	several values of $T$. In the QPM
the bulk corresponds to a bath of quasi-particles, that is, quarks and
gluons, with temperature-dependent masses.
	All the results have been obtained by solving
	Eq.~\eqref{eq:L1bis3} hence taking both diffusion and drag into account.

\subsection{Momentum randomization\label{sec:equi}}

	\begin{figure}[t!]
		\begin{center}
			\includegraphics[scale=0.23]{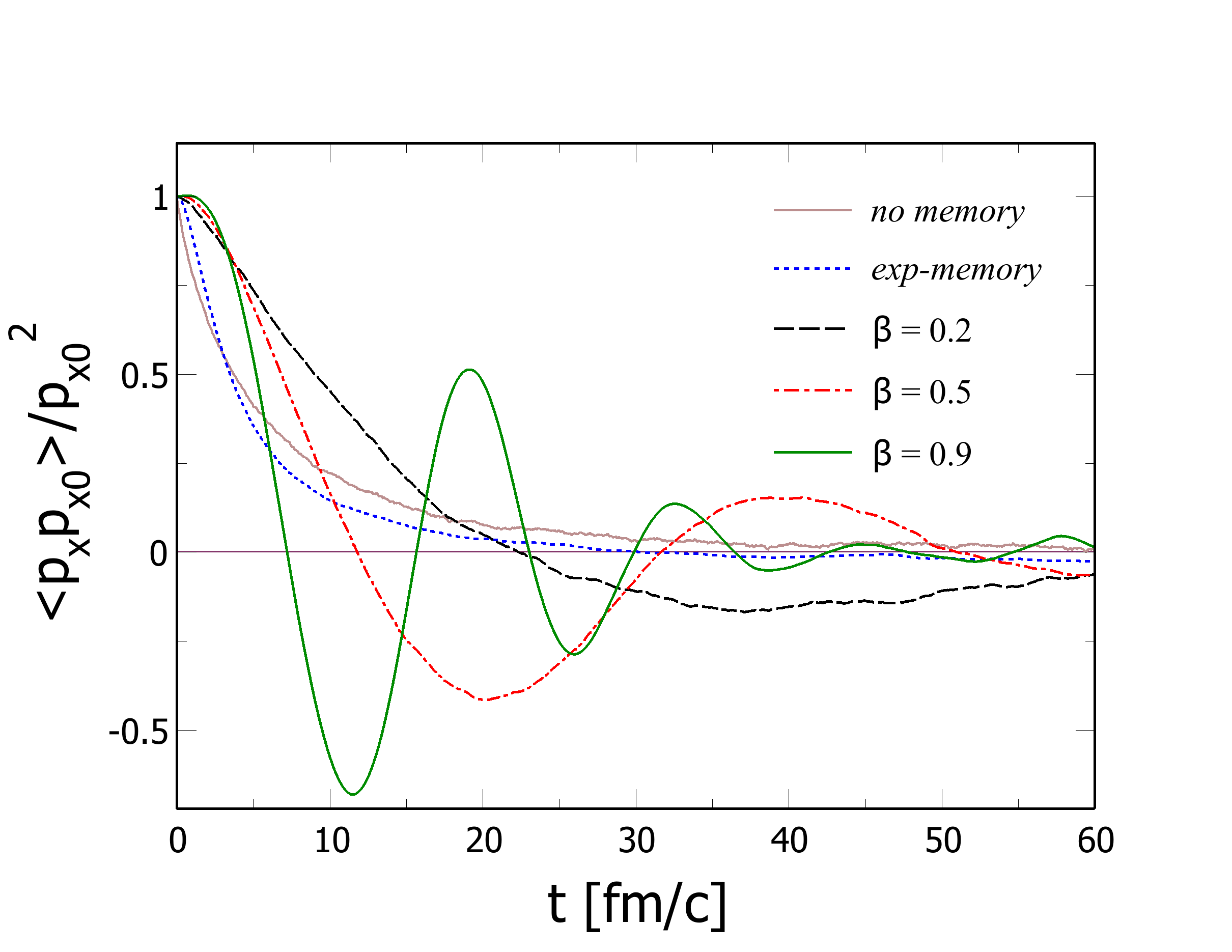}
		\end{center}
		\caption{\label{Fig:pxd02}
			$\langle p_x\rangle$
			versus time for three values of $\beta$,
			$\tau=1$ fm/c, $p_0=1$ GeV,
			at $T=1$ GeV and $\mathcal{D}=0.5$ GeV$^2$/fm.
			For the exp-memory calculation we used
			$\tau=1$ fm/c.
			}
	\end{figure} 
	
	In the upper panel of
	Fig.~\ref{Fig:pxd02} we plot $\langle p_x p_{x0}\rangle$ in units
	of $p_{x0}^2$, where $p_{x0}$ denotes the initial value of $p_x$,
	versus time for three values of $\beta$, for $\tau=1$ fm/c,
	at $T=1$ GeV and $\mathcal{D}=0.5$ GeV$^2$/fm;
	the results for $\langle p_y p_{y0}\rangle$ and
	$\langle p_z p_{z0}\rangle$ are similar.
	The value of $\mathcal{D}$ is about
	the magnitude of the average 
	diffusion coefficient derived from pQCD
	at the same temperature, 
	and rescaled by the $k-$factor that generates
	$R_\mathrm{AA}(p_T)$ which agree with experimental
	data in realistic simulations~\cite{Scardina:2017ipo}.
	In the non-relativistic limit, $\langle p_x p_{x0}\rangle$
	is proportional to the autocorrelation function of velocity
	which is largely studied in models of stochastic processes 
	with memory kernels.    
	$\langle p_x p_{x0}\rangle$ allows us to study how correlations
	of momentum with the initial condition are washed out by the 
	interactions of the particle with the bath.
	For comparison, in this figure we also show one result obtained
	assuming an exponential memory kernel that follows the
	implementation of~\cite{Ruggieri:2022kxv},
	with a memory time $\tau=1$ fm/c and the same value of
	$\mathcal{D}$.
	In Fig.~\ref{Fig:pxd02} we also show the result for a
	the calculation for a bath without memory,
	characterized also by an exponential decay.
	We note that 
	for the noise with the exponential kernel
	the behavior of the correlator follows 
	that of the memoryless process, besides a small
	delay in the very early stage. On the other hand,
	the behavior of the correlator 
	for the bath with the noise
	in Eq.~\eqref{eq:Caputo_111} 
	is somehow different. 
	For the small $\beta=0.2$, we find no big qualitative
	difference between the memoryless and the exponential
	cases, besides some delay of the momentum randomization:
	this is not very surprising since we already discussed in the previous section that for $\beta\rightarrow0$,
	the correlations of the noise decay quickly.
	For larger $\beta$, $\langle p_x p_{x0}\rangle$ 
	develops oscillations, signaling that the randomization
	of momentum is nontrivial. It is likely that these oscillations
	are related to a continuous energy exchange between the bath and
	the HQ, as it becomes evident from the results on the kinetic energy that we show later.
	The qualitative
	behavior of $\langle p_x p_{x0}\rangle$ that we found
	is in agreement with previous model calculations 
	of the velocity autocorrelation function in the non-relativistic
	limit~\cite{kneller}:
	in the latter reference, 
	a different memory kernel was used, nevertheless the trend
	of the correlator is similar in the two calculations,
	including the fact that enhancing the correlations
	of the noise results in wider fluctuations of 
	$\langle p_x p_{x0}\rangle$.

\subsection{Thermalization for a simple initialization\label{sec:thermalizaaa}}	
	
	Momentum randomization is not enough to make
	statements about
	thermalization: in fact, one should also check that
	the kinetic energy per particle
	of the HQs corresponds to the average value
	expected from a thermal distribution at that temperature,
	$K_\mathrm{eq}$, given by 
	\begin{equation}
		K_\mathrm{eq}
		\equiv
		\frac{\int\frac{d^3p}{(2\pi)^3}
			( \sqrt{p^2 + m^2}-m )e^{-\sqrt{p^2+m^2}/T}}
		{\int\frac{d^3p}{(2\pi)^3}e^{-\sqrt{p^2+m^2}/T}}.
		\label{eq:eq_1}
	\end{equation}
$K_\mathrm{eq}$ can be computed analytically for any value of
$T$ and $m$ by using standard integral representations
of the modified Bessel functions,
that lead at
\begin{equation}
	K_\mathrm{eq} =3T\left[1 -\frac{m}{3T} + 
	\frac{m}{3T}\frac{K_1(m/T)}{K_2(m/T)}
	\right].
	\label{eq:eq_anal_222}
\end{equation}
In the non-relativistic limit $K_\mathrm{eq}  =3T/2$, while in the
ultrarelativistic case $K_\mathrm{eq}  =3T$.

		\begin{figure}[t!]
		\begin{center}
			\includegraphics[scale=0.23]{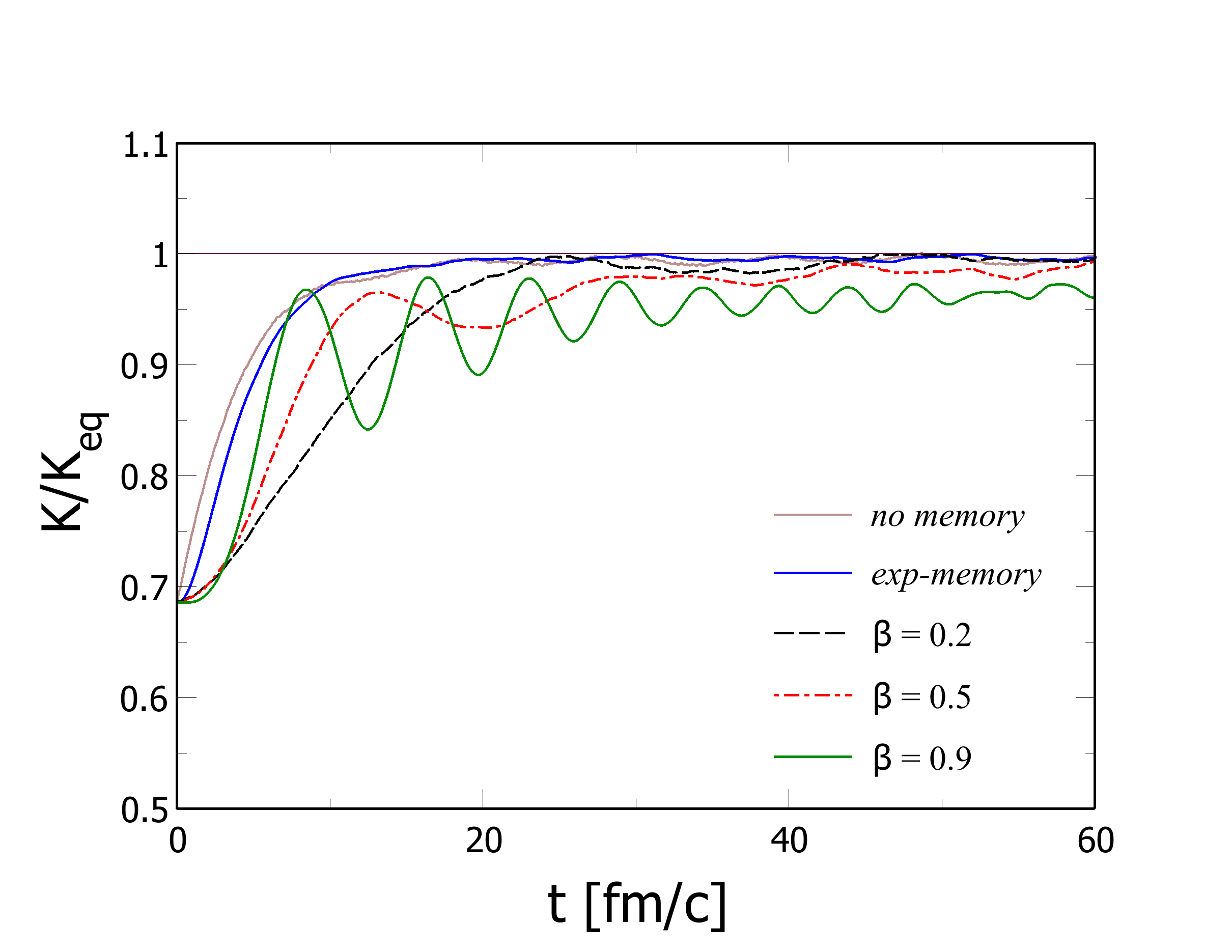}
		\end{center}
		\caption{\label{Fig:epsp05p1}$K/K_\mathrm{eq}$ 
			versus time for 
			$\tau=1$ fm/c, 
			at $T=0.25$ GeV and $\mathcal{D}=0.1$ GeV$^2$/fm. 
			Initialization corresponds to  $p_T=1$ GeV and $p_z=0$.
			The value of $\mathcal{D}$ was chosen in agreement
			with the diffusion coefficient computed within the
			QPM at the same temperature and $p=1$ GeV.
	$K_\mathrm{eq}=0.44$ GeV.
		}
	\end{figure} 

In this subsection, 
in order to emphasize the qualitative effects of memory on
thermalization,
we analyze a simple initialization
corresponding to $p_T=1$ GeV and $p_z=0$:
the latter corresponds to the midrapidity region
of realistic collisions.
In Fig.~\ref{Fig:epsp05p1} 
we plot $K \equiv\langle \sqrt{p^2 + m^2}-m  \rangle$
in units of $K_\mathrm{eq}$ versus time obtained by our calculations for three values of $\beta$, $\tau=1$ fm/c, at $T=0.25$ GeV and $\mathcal{D}=0.1$ GeV$^2$/fm, 
in agreement with the coefficient 
computed within the QPM at $p=1$ GeV at the same temperature.
Moreover,
$K_\mathrm{eq}\approx 0.44$ GeV
corresponds to the equilibrium value for 
	$T=0.25$ GeV
	and $m=1.5$ GeV.
	For comparison, we also plot $K/K_\mathrm{eq}$ for a memoryless
	process and for a process with an exponentially decaying memory,
	as we did in Fig.~\ref{Fig:pxd02}.

	We note that
	for the two smaller values of $\beta$
	in Fig.~\ref{Fig:epsp05p1}, the average kinetic energy
	approaches $K_\mathrm{eq}$ within the
	time range explored, meaning that the HQs eventually thermalize
	with the medium; we also note that increasing $\beta$
	from $0.2$ to $0.5$ results in 
	a few oscillations of $K/K_\mathrm{eq}$. 
	On the other hand, we note that
	for $\beta=0.9$ the average kinetic energy of the HQs
	remains smaller than $ K_\mathrm{eq}$, 
	meaning that in this case
	thermalization is not complete, due to the correlations of the
	noise. In contrast, the memoryless and the exponential 
	bath lead eventually to thermalization.
We leave the estimate of the thermalization time to the case
of realistic initialization in the next subsection:
here it is enough to remark that
our results
	suggest that memory in the bath results in the
	slowing down of thermalization of HQs.

\subsection{Thermalization time for realistic initializations\label{sec:impactonraa}}

In the previous subsection we illustrated the effects
of memory on thermalization for charm quarks, using
a simplified initialization.
In this subsection, we quantitatively
study the thermalization time,
$\tau_\mathrm{therm}$, of
charm and beauty, firstly focusing on its dependence on $\mathcal{D}$. 
Differently from the previous subsections,
here we initialize HQs by means of the realistic Fixed Order + Next-to-Leading Log (FONLL) distribution~\cite{FONLL,Cacciari:2012ny},
\begin{equation}
\left(\frac{dN}{d^2p_T}\right)_\mathrm{FONLL} =
\frac{x_0}{(1+x_3 p_T^{x_1})^{x_2}},
\label{eq:ffnlo_123}
\end{equation}	
where the parameters are $x_0=20.2837$, $x_1=1.95061$,
$x_2=3.13695$, $x_3=0.075166$ for charm and
$x_0=0.46799$, $x_1=1.83805$,
$x_2=3.07569$, $x_3=0.030156$
for beauty. When the bath has no memory,
the thermalization time is estimated by studying the
decay of one component of the momentum of HQs: in this case
$p_x=p_0 e^{-\gamma t}$ with $\tau_\mathrm{therm}=1/\gamma$.
see Fig.~\ref{Fig:pxd02}.
When the bath has memory this definition of $\tau_\mathrm{therm}$
does not seem to be appropriate, because the average of the
components of $p$ fluctuate,
see again Fig.~\ref{Fig:pxd02}. 
Furthermore, more generally we have already seen that
thermalization is delayed with respect to the memoryles
case for which $\tau_\mathrm{therm}=1/\gamma$.

In order to estimate
$\tau_\mathrm{therm}$ in this case, we proceed as follows.
We firstly compute $\langle p_T\rangle$
of HQs corresponding to the initialization in
Eq.~\eqref{eq:ffnlo_123}, and
assume $p_z=0$ to mimick the midrapidity region of the collisions.
Then, 
we prepare an initialization with $p_{x0}=\langle p_T \rangle$,
$p_{y0}=0$.
For a given $\mathcal{D}$ and $T$,
we compute $\tau_\mathrm{therm}$
for the bath without memory by 
fitting $\langle p_x\rangle$ with an exponential function
$p_x=p_{x0} e^{-t/\tau_\mathrm{therm}}$,
then we compute $\Upsilon_\mathrm{therm}\equiv K/K_\mathrm{eq}$ at $t=\tau_\mathrm{therm}$;
we repeat the procedure for other values of $\mathcal{D}$
at the same $T$. We find that the value of $\Upsilon_\mathrm{therm}$
is not very sensitive to the value of $\mathcal{D}$,
and that $\tau_\mathrm{therm}\approx 1/\gamma=\langle E\rangle T/\mathcal{D}$ where $\langle E\rangle$ denotes the initial
average energy of HQs from the distribution~\eqref{eq:ffnlo_123}.
We then use $\Upsilon_\mathrm{therm}$ at that $T$
to 
estimate $\tau_\mathrm{eq}$ for the bath with memory for each value
of $\mathcal{D}$,
by identifying in the latter case the thermalization time with
the time at which $K/K_\mathrm{eq}=\Upsilon_\mathrm{therm}$.

	\begin{figure}[t!]
	\begin{center}
		\includegraphics[scale=0.23]{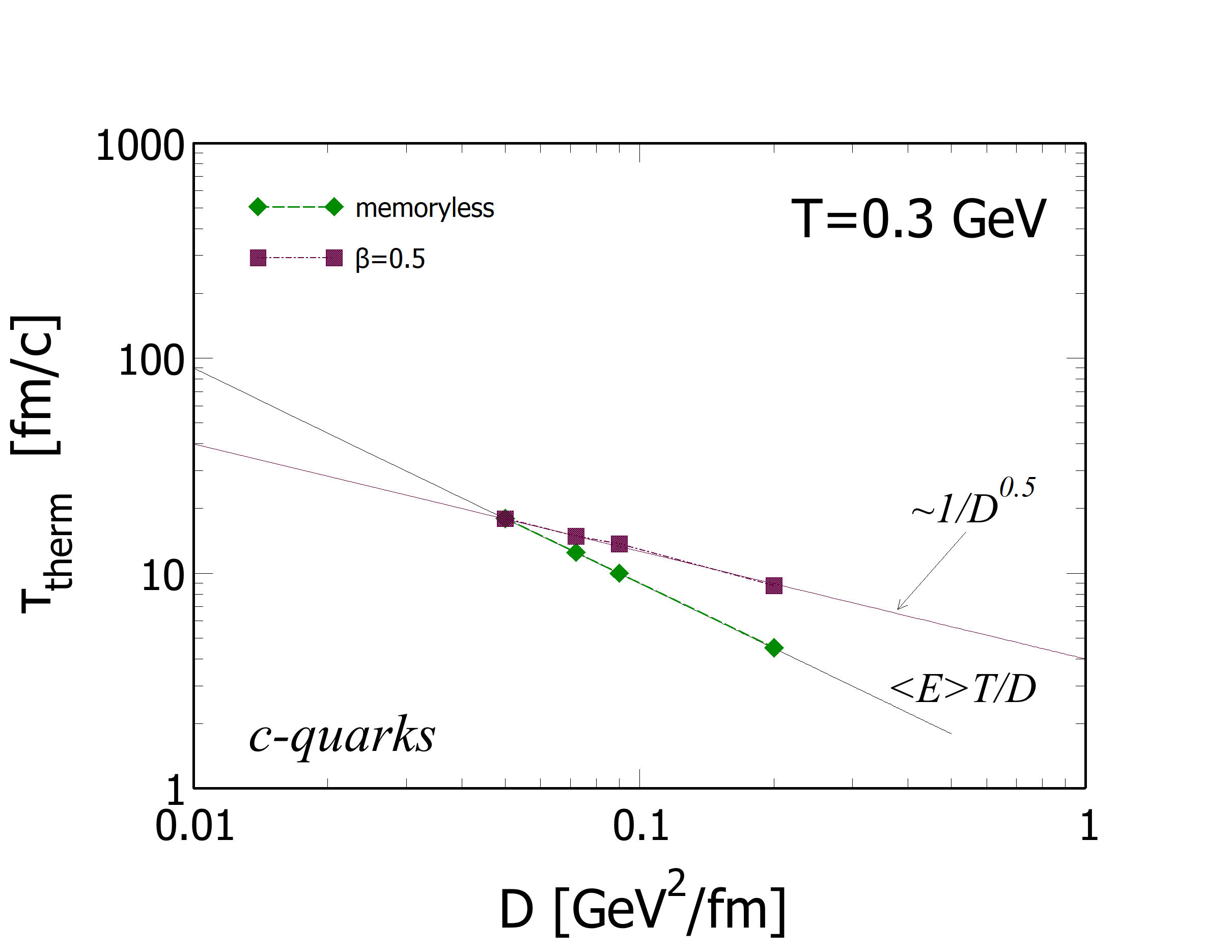}
	\end{center}
	\caption{\label{Fig:dc0305}$\tau_\mathrm{therm}$
		versus $\mathcal{D}$ for charm quarks at $T=0.3$ GeV.
		Green diamonds correspond to the bath without memory
		while maroon squares to that with a power law memory
		with $\beta=0.5$. 
	}
\end{figure}

In Fig.~\ref{Fig:dc0305} we plot 
$\tau_\mathrm{therm}$
versus $\mathcal{D}$ for charm quarks at $T=0.3$ GeV;
we checked that the behavior is qualitatively similar for
other temperatures as well as for beauty quarks.
For the bath with memory we show the data for the case 
$\beta=0.5$ only, since the other cases are qualitatively similar.
As already mentioned, for the bath without memory we find 
$\tau_\mathrm{therm}\approx \langle E\rangle T/\mathcal{D}$, 
in agreement with the Fluctuation-Dissipation
Theorem.
For $\beta=0.5$ we find 
$\tau_\mathrm{therm}\propto 1/\mathcal{D}^\alpha$
with $\alpha\approx 0.5$.  
In order to quantify the effect of 
memory on $\tau_\mathrm{therm}$,
we consider
the phenomenologically relevant $\mathcal{D}$ within
the QPM 
model at $p=0$, namely
this gives $\mathcal{D}=0.13$ GeV$^2$/fm at $T=0.3$ GeV
( k-factors included).
With this value of $\mathcal{D}$, for the bath
without memory we get
$\tau_\mathrm{therm}=6.9$ fm/c at $T=0.3$ GeV,
while for the bath with memory we get 
$\tau_\mathrm{therm}=11.4$ fm/c at the same temperature.

\begin{figure}[t!]
	\begin{center}
		\includegraphics[scale=0.23]{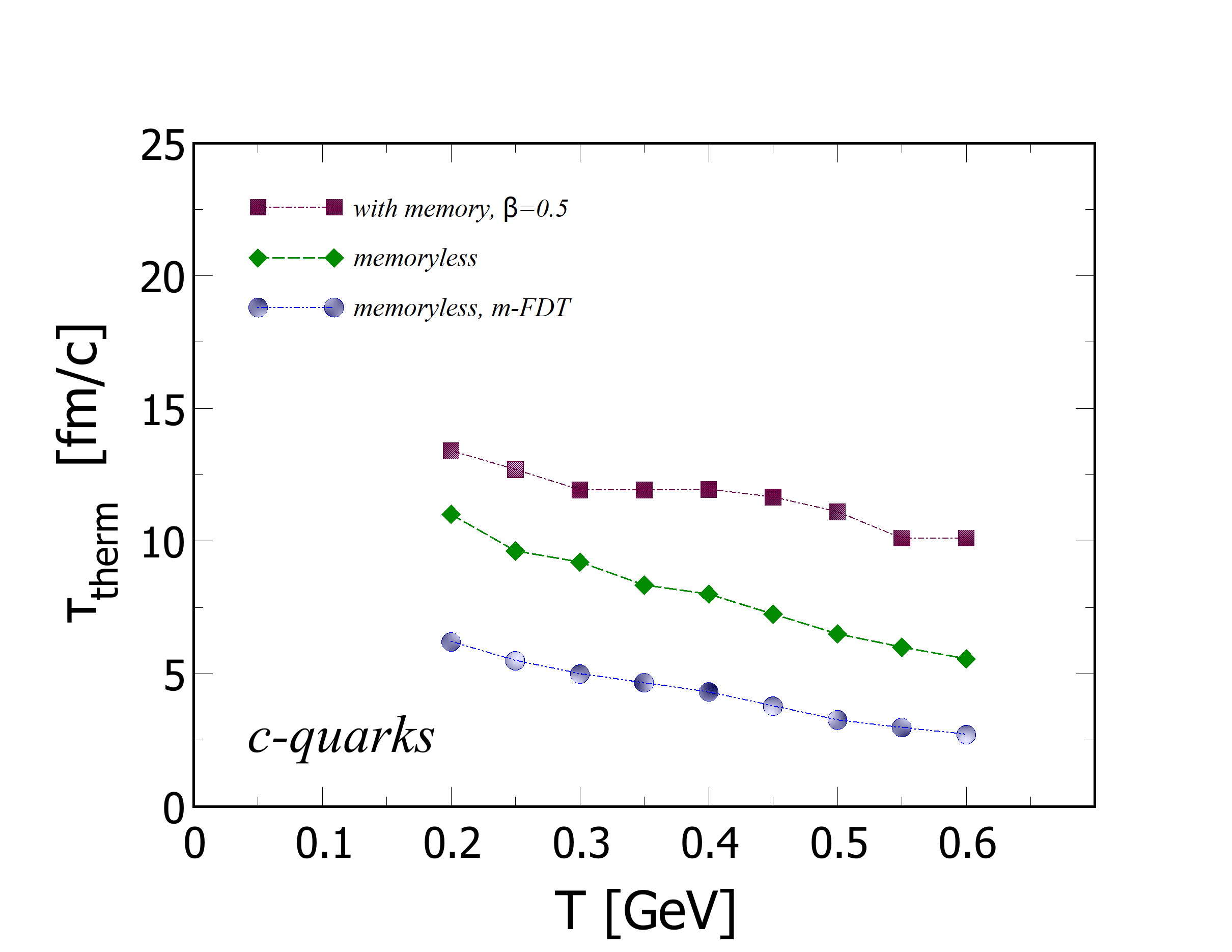}\\
		\includegraphics[scale=0.23]{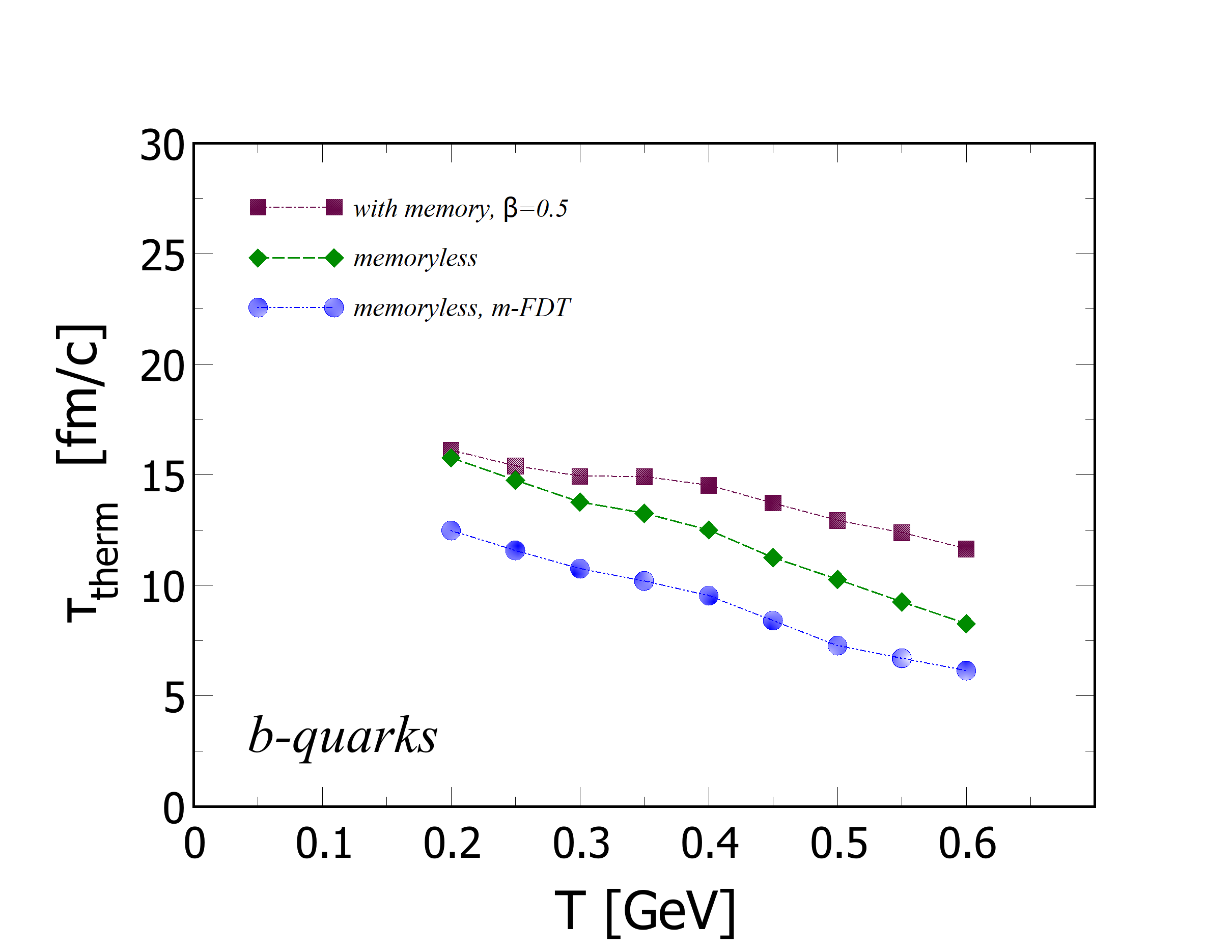}
	\end{center}
	\caption{\label{Fig:tctb}$\tau_\mathrm{therm}$
		versus $T$ for charm
		(upper panel) and beauty (lower panel). 
	Maroon squares correspond to the thermalization time 
	obtained with memory for $\beta=0.5$.
	Green diamonds denote the memoryless thermalization time
	computed by fitting
	$p_x(t)$ by $p_x=p_0 e^{-t/\tau_\mathrm{therm}}$.
	Finally,
blue circles correspond to $\tau_\mathrm{therm}=1/\gamma$,
with $\gamma$ computed by the Eq.~\eqref{eq:FDT_in_h_mc}
	in which we replaced $E$ by $m$.
	}
\end{figure} 
 
In Fig.~\ref{Fig:tctb}
we plot
$\tau_\mathrm{therm}$
versus $T$ for charm
(upper panel) and beauty (lower panel). 
At each temperature, the diffusion coefficient is that
of the QPM model computed at $p=0$.
For the sake of comparison, we also plot 
the
blue circles corresponding  to 
\begin{equation}
\tau_\mathrm{therm}=
\frac{m T}{\mathcal{D}},
\label{eq:deviat}
\end{equation}
that amounts to replace $E$ by $m$ in the FDT~\eqref{eq:FDT_in_h_mc}.
We note that already taking into account the initial average
kinetic energy of the HQs amounts to an increase of 
$\tau_\mathrm{therm}$ in comparison with the result that we would
get if we defined $\tau_\mathrm{therm}$ by virtue of Eq.~\eqref{eq:deviat}.
Finally, the maroon squares correspond to $\tau_\mathrm{therm}$
for the medium with memory; for the sake of concreteness we
only show the results for $\beta=0.5$.
In this case, $\tau_\mathrm{therm}$ was defined by comparing $K/K_\mathrm{eq}$ in the cases with and without memory
as explained above. We note that the effect of memory is to
increase the thermalization time of HQs;
the difference between the cases with and without memory
increase with temperature, and is milder for beauty quarks.

\begin{figure}[t!]
	\begin{center}
		\includegraphics[scale=0.23]{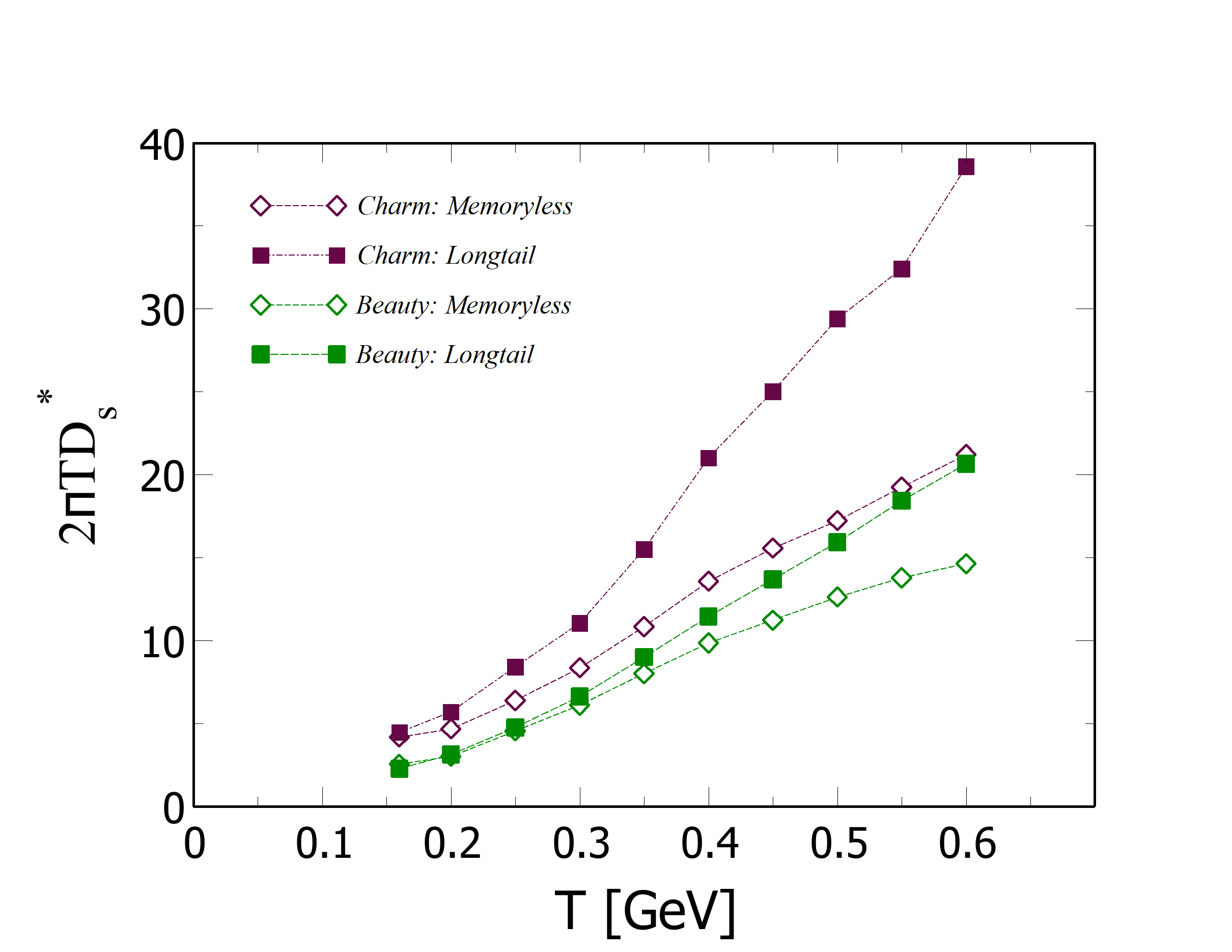}
	\end{center}
	\caption{\label{Fig:2ptds}$2\pi T D_s^*$
		versus $T$ for charm and
	beauty quarks. Open symbols correspond to 
	calculations for the memoryless bulk,
solid symbols represent calculations for the bulk with 
longtail memory
with $\beta=0.5$.
	}
\end{figure} 

We can define an effective spatial diffusion coefficient, $D_s^*$,
by virtue of the relation
\begin{equation}
D_s^* = \frac{T}{\langle E \rangle }\tau_\mathrm{therm},
\label{eq:tauds2}
\end{equation}
where $\langle E \rangle$ denotes the initial average energy
of the HQs;
the definition~\eqref{eq:tauds2} 
gives back the commonly used $D_s=T^3/\mathcal{D}$
when $\tau_\mathrm{therm}=mT/\mathcal{D}$ and $\langle E \rangle$
is replaced by $m$.
In Fig.~\ref{Fig:2ptds} we plot $2\pi T D_s^*$
versus $T$ for charm and
beauty quarks. Results are shown for the memoryless case
as well as for the memory case.
We note that the memory in the bulk 
leads at the increase of $D_s^*$; the effect is more important
for the charm quarks, and the discrepancy between the results
obtained with and without memory increases with temperature. 

Summarizing, the meaning of the results collected in
Fig.~\ref{Fig:2ptds} is that due to memory effects,
the system appears to have a larger $D_s^*$.
It can delays the formation of the $R_\mathrm{AA}(p_T)$ with memory.
To reproduce the same $R_\mathrm{AA}(p_T)$ with memory as of the case without memory, one needs to reduce the magnitude of $D_s^*$.
Consequently, if one tries to get the spatial diffusion
coefficients that fit the
experimental data discarding the memory, then gets
larger values of $D_s$ with respect to the real ones.
However, it is interesting that for beauty quarks this effect
appears to be negligible between $T_c$ and $3T_c$ even considering
strong memory $\beta=0.5$. 
This is another feature that makes the extraction of $D_s(T)$
for beauty quarks more solid (and directly comparable
to lattice QCD).

\subsection{Momentum isotropization\label{subsec:momisoaaa}}

	\begin{figure}[t!]
	\begin{center}
		\includegraphics[scale=0.23]{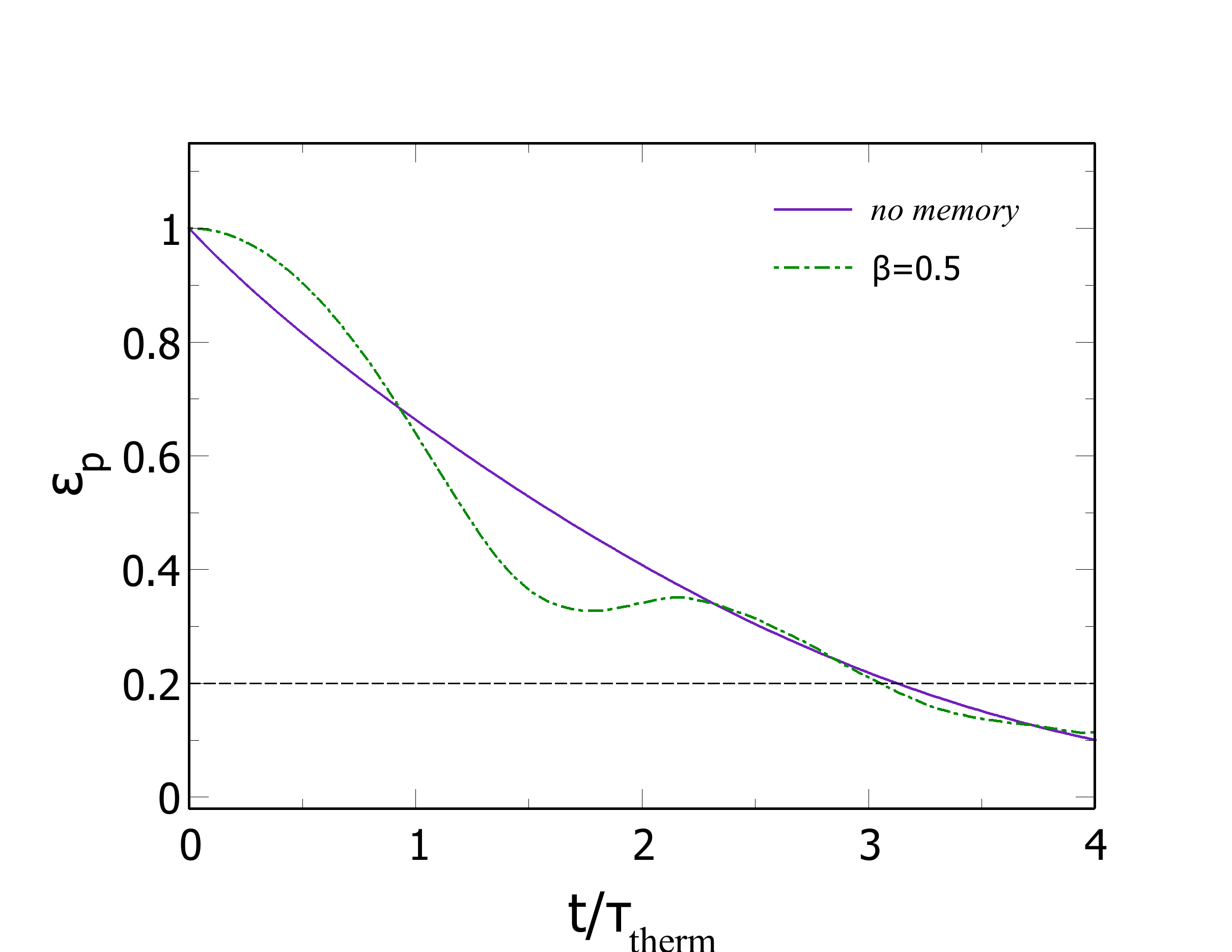}
	\end{center}
	\caption{\label{Fig:epsp05p1mm}	$\varepsilon_p$ 
		versus $t/\tau_\mathrm{therm}$,
		at $T=0.3$ GeV and $\mathcal{D}=0.09$ GeV$^2$/fm. 
		Initialization corresponds to  the FONLL distribution 
		and $p_z=0$.
		The value of $\mathcal{D}$ was chosen in agreement
		with the diffusion coefficient computed within the
		QPM at the same temperature and $p=0$ GeV.
	}
\end{figure}

We close this study by briefly analyzing the
isotropization of the heavy quarks.  
In the midrapidity region of realistic collisions, HQs 
are produced with a finite average $p_T$ and $p_z\approx0$:
the initial condition is therefore anisotropic in momentum, 
but the interaction of the HQs with the bath might lead to
momentum isotropization.
We quantify the momentum anisotropy by introducing
an eccentricity, $\varepsilon_p$, as
\begin{equation}
	\varepsilon_p=\frac{\langle p_T^2-2p_z^2\rangle}
	{\langle p^2_T + p_z^2 \rangle}.
	\label{eq:epsp_def}
\end{equation}
In Fig.~\ref{Fig:epsp05p1mm}, 
we plot $\varepsilon_p$
versus time for the memoryless process and for the
power law memory process with $\beta=0.5$.
Initialization corresponds to the FONLL distribution with 
$p_z=0$. 
Time is measured in units of the thermalization time for the two cases,
which corresponds to $\tau_\mathrm{therm}=8.3$ fm/c
for the memoryless
case and $\tau_\mathrm{therm}=11.76$ fm/c
for the power law memory case.
Moreover, we used $T=0.3$ GeV and $\mathcal{D}=0.09$ GeV$^2$/fm corresponding to
the diffusion coefficient computed within the
QPM at the same temperature and $p=0$ GeV,
in agreement with the value used in the previous subsection
to compute the thermalization time.
We checked that for other values of $\beta$
$\varepsilon_p$ qualitatively behaves similarly.

Initially
$\varepsilon_p=1$ since $p_z=0$; 
however, interactions with the medium 
lead to $\varepsilon_p\rightarrow 0$ namely to
momentum isotropization.
Assuming that a fair amount of isotropization
takes place when $\varepsilon_p=0.2$, 
we find that isotropization time, $\tau_\mathrm{iso}$,
is  $\tau_\mathrm{iso}\approx 2.54\tau_\mathrm{therm}$ 
for the memoryless
case, while $\tau_\mathrm{iso}\approx 2.76\tau_\mathrm{therm}$  for the $\beta=0.5$ case.
We conclude that although memory delays 
both thermalization and isotropization,
$\tau_\mathrm{therm}$ and $\tau_\mathrm{iso}$ lie
in the same ballpark.

	%%%%%%%%%%%%%%%%%%%%%%%%% Start Conclusions %%%%%%%%%%%%%%%%%%%%%%%%%%%%%%%%
	\section{Conclusions and outlook\label{sec:conclusions}}
	
	We studied the effects of a power law
	correlated noise 
	on momentum randomization, isotropization
	and thermalization of heavy quarks (HQs),
	in a thermal bath.	
	Our work is related to the problem of HQs in relativistic nuclear collisions, in which HQs themselves diffuse and lose energy in the quark-gluon plasma (QGP), as well as in
	the very early stage in which the dynamics 
	of the bulk is dominated
	by a dense gluon system.
	This work is a 
	follow-up of~\cite{Ruggieri:2022kxv} in which
	the same problem was studied only with the
exponential correlator of the random force:
	in the present work, we focused on a power law correlator.
	The noise with the desired correlations, $h$,
	was generated by a convenient superposition
	of gaussian white noises, see Eq.~\eqref{eq:Caputo_111}.
	In this definition, two parameters enter:
	$\tau$, the memory time, that sets the time scale
	at which correlations decay, see for example
	Eq.~\eqref{eq:Caputo_corr_111_asy_2}, 
	and $\beta$, which changes the power law
	of the decay of the correlator,
	see again Eq.~\eqref{eq:Caputo_corr_111_asy_2}:
	increasing $\beta$ from $0$ to $1$ results in
	the slower decay of the correlator at large times,
	hence in some sense, increasing $\beta$ while
	keeping $\tau$ fixed amounts to have a bath
	with more persistent memory.
	The interaction of the heavy quarks with the bath
	at a fixed temperature
	were modeled by a generalized Langevin equation,
	in which the random force, $\eta$, is assumed
	to be time-correlated and the dissipative kernel is
	defined by a Fluctuation-Dissipation-Theorem-like 
	(FDTlike) equation.

	We studied momentum randomization, thermalization and
	momentum isotropization of HQs
	by using the whole Langevin equation~\eqref{eq:L1}.
	Initializing HQs with the particular initial
	condition $p_x\neq 0$, $p_y=p_z=0$,
	we found that 
	the qualitative behavior of $\langle p_x\rangle$
	with the power law memory can be quite different from the
	memoryless case: in fact, 
	the exponential decay expected in the latter case is replaced
	by damped oscillations in the former case,
	and the oscillations become more persistent by 
	increasing $\beta$.
	This is in agreement with our
	general understanding, since increasing $\beta$
	results in injecting more correlations in the
	random force, hence an HQ needs more time to forget
	about its initial condition. Our results show that momentum
	randomization is not a trivial process 
	when HQs interact with a medium with
	power law memory.
	
	We found that memory slows down thermalization and
	momentum isotropization of the HQs. 
	Thermalization times are increased by memory:
	this leads to the increase of the spatial diffusion coefficient.
	We found that the effect on charm quarks is substantial,
	while that on beauty quarks is smaller. This is probably
	due to the fact that in general the thermalization time
	of beauty quarks are larger than those of charms,
	hence  for the former the correlations of the noise
	have enough time to decay before thermalization sets in.
	We also found that momentum isotropization 
	is slightly delayed by memory; however,
	isotropization times
	starting with the 
	FONLL distribution are in the same ballpark of the
	thermalization ones.
	
	In the memoryless case the thermalization time,
	$\tau_\mathrm{therm}$, is $\propto 1/\mathcal{D}$;
on the other hand, in the case of the non-Markovian dynamics,
the dependence of $\tau_\mathrm{therm}$ on $\mathcal{D}$ is 
different. In particular, for $\beta=0.5$ we found 
$\tau_\mathrm{therm}\propto 1/\mathcal{D}^{0.5}$.
This leads to the increase of $\tau_\mathrm{therm}$ due to
memory, which we quantified to be of the order of $30\%$ for the
charm quarks for temperatures between $T_c$ and $3T_c$.
Interestingly, the impact of memory 
on $\tau_\mathrm{therm}$ of beauty
quarks is damped in the same range of temperatures
in the QGP phase.

This work paves the way to more realistic implementations,
that should include a proper initial geometry as well as
an expanding medium.
In the future, it will be relevant to investigate if the
relation between $R_\mathrm{AA}$ and $v_2$ is modified
by a non-Markovian dynamics, and to quantify the potential
effects on observables for both charm and beauty quarks. 
We believe that memory has a potential effect on  observables.
Following~\cite{Ruggieri:2022kxv} we expect that 
memory delays the formation of the $R_\mathrm{AA}(p_T)$;
consequently, given a diffusion coefficient and a time interval,
memory will keep the values of  $R_\mathrm{AA}(p_T)$ higher.
This implies that one would need a higher 
momentum diffusion coefficient $\mathcal{D}$, 
or equivalently, a smaller $D_s$, in order to
reproduce the experimental $R_\mathrm{AA}(p_T)$.
In other words, neglecting memory and fixing the 
transport coefficients in order to reproduce data
would lead to an overestimate of $D_s$.
We aim at discussing in detail these problems in future publications.

	%%%%%%%%%%%%%%%%%%%%%%%%% End Conclusions %%%%%%%%%%%%%%%%%%%%%%%%%%%%%%%%

% !!!!!!!!!!!!!!!!!! Start : Acknowledgements !!!!!!!!!!!!!!!!!! %
	\begin{acknowledgements}
		M.R. acknowledges John Petrucci for inspiration,
and
V. Minissale, G. Nugara, G. Parisi, S. Plumari,  M. L. Sambataro
and in particular L. Oliva for the numerous discussions on the topics
presented in this article.
		S.K.D. acknowledges the support from DAE-BRNS, India, Project No. 57/14/02/2021-BRNS. V. G. acknowledges the support from
		HQCDyn Linea 2 UniCT.
	\end{acknowledgements}
	% !!!!!!!!!!!!!!!!!! End : Acknowledgements !!!!!!!!!!!!!!!!!! %

	% xxxxxxxxxxxxxxxxxxxxxxxxxxxxxxxxxxxxxxxxxxxxxxxxxxxxxxxxxxxxxxxxxxxxxxxxxx    

\end{document}